\documentclass[prb,aps,twocolumn,floats,amsmath,amssymb,preprintnumbers]{revtex4}
\usepackage{graphicx}
\usepackage{dcolumn}
\usepackage{bm}

\begin{document}
\preprint{(submitted to J. Chem. Phys.)}

\title{Nature of Ar bonding to small Co$_n^+$ clusters and its effect\\ on the structure determination by far-infrared absorption spectroscopy}

\author{Ralf Gehrke}
\author{Philipp Gruene}
\author{Andr\'e Fielicke}
\author{Gerard Meijer}
\author{Karsten Reuter}

\affiliation{Fritz-Haber-Institut der Max-Planck-Gesellschaft,
Faradayweg 4-6, D-14195 Berlin, Germany}

\received{13th October 2008}

\begin{abstract}
Far-infrared vibrational spectroscopy by multiple photon dissociation has proven to be a very useful technique for the structural fingerprinting of small metal clusters. Contrary to previous studies on cationic V, Nb and Ta clusters, measured vibrational spectra of small cationic cobalt clusters show a strong dependence on the number of adsorbed Ar probe atoms, which increases with decreasing cluster size. Focusing on the series Co$_4^+$ to Co$_8^+$ we therefore use density-functional theory to analyze the nature of the Ar-Co$_n^+$ bond and its role for the vibrational spectra. In a first step, energetically low-lying isomer structures are identified through first-principles basin-hopping sampling runs and their vibrational spectra computed for a varying number of adsorbed Ar atoms. A comparison of these fingerprints with the experimental data enables in some cases a unique assignment of the cluster structure. Independent of the specific low-lying isomer, we obtain a pronounced increase of the Ar binding energy for the smallest cluster sizes, which correlates nicely with the observed increased influence of the Ar probe atoms on the IR spectra. Further analysis of the electronic structure motivates a simple electrostatic picture that not only explains this binding energy trend, but also why the influence of the rare-gas atom is much stronger than in the previously studied systems.
\end{abstract}

\maketitle

\section{Introduction}

The intricate relationship between structural and electronic degrees of
freedom of metal clusters in the sub-nm size range dictates a precise
knowledge of the geometric structure when aiming to exploit the unique
optical, magnetic and chemical properties of these materials. Experimentally,
obtaining information on the atomic arrangement of such small clusters in the
gas phase remains a challenge. Investigations of the reactivity of
metal clusters with small molecules, in particular their saturation behavior,
have been used to draw conclusions about the clusters' atomic
arrangements.\cite{parks94} Photoelectron spectroscopy provides direct
information on the electronic structure of clusters which can be related to
their geometries.\cite{li03} Information on the shape of cluster ions can be
obtained from ion mobility measurements\cite{helden93,ho98,weis05} and,
recently, electron diffraction of trapped charged clusters has shown to be
promising for revealing their structure.\cite{schooss05,blom06,xing06}
Another possibility that is particularly sensitive to the internal cluster
structure is the measurement of vibrational frequencies. A corresponding
technique that has recently been successfully employed to determine the
structure of cationic and neutral metal clusters containing between three and
20 atoms is far-infrared (vibrational) resonance enhanced multiple photon
dissociation (FIR-MPD)
spectroscopy.\cite{fielicke04,ratsch05,fielicke05,fielicke07,gruene07} The
essential idea of this technique is to irradiate rare-gas complexes of the
targeted clusters with infrared (IR) light in the range of the
structure-specific vibrational fundamentals. When the IR light is resonant
with a vibrational mode in the cluster complex, the complex can absorb
several photons and subsequently evaporate off one or more rare-gas atoms.
Recording the resulting abundance changes of the rare-gas complexes as a
function of the IR frequency yields the desired spectra that can then be
compared to computed IR absorption spectra for different isomer structures
obtained e.g. using density-functional theory (DFT). As long as different
isomers exhibit distinct vibrational fingerprint patterns this then enables a
unique determination of the cluster structure.

A fundamental
assumption behind this approach is that the rare-gas atoms do not
significantly influence the vibrational spectrum and merely act as a probe
for detecting the resonant absorption of IR photons by the metal
cluster.\cite{fielicke05a} A negligible influence of the rare-gas
atoms was indeed inferred from insignificant differences in the
IR-spectra of previously studied cationic V$_n^+$ ($n=3-23$)
\cite{fielicke04,ratsch05}, Nb$_n^+$ ($n=5-9$) \cite{fielicke05,fielicke07}
and Ta$_n^+$ ($n=6-20$) \cite{gruene07} complexes that contained one or more
Ar atoms. Restricted test calculations comparing the IR-spectra of bare and
Ar-complexed V$_3^+$ and V$_4^+$ clusters led to the same conclusion
\cite{fielicke04,ratsch05}, so that the theoretical modeling for the
mentioned systems focused exclusively on the IR spectra of the bare metal
clusters.

We find the situation to be markedly different for small
cationic Co clusters. Here, the measured FIR-MPD spectra show an intriguing
dependence on the number of adsorbed Ar atoms, which is strongest for
clusters containing less than six cobalt atoms. This prompts us to perform a
systematic DFT analysis of the nature of the Ar-Co$^+_n$ bond and its
implications for the vibrational spectra. For the series Co$^+_4$ to Co$^+_8$
we first employ basin-hopping (BH) \cite{wales00} to sample the
configurational space and identify energetically low-lying isomers.
Explicitly accounting for the Ar probe atoms, a comparison of the computed IR
spectra for these isomers with the measured data allows in some cases for a
unique assignment of the experimental cluster structure. The calculations
reveal furthermore a characteristic increase of the Ar binding energy for the
smaller clusters, rationalizing the observed increased influence of the Ar
probe atoms on the IR spectra. We can trace this binding energy trend back to
the predominant contribution to the Ar-Co$^+_n$ bond arising from the
polarization of the rare-gas atom in the electrostatic field of the cationic
cluster. This motivates a simple electrostatic model that not only reproduces
the binding energy trend, but also explains why the interaction of Ar is much
stronger than in the previously studied systems, where little influence of
the probe atoms on the vibrational spectra is observed.

\section{Experiment}

The experiments are carried out in a molecular-beam setup that is coupled to
a beam line of the Free Electron Laser for Infrared eXperiments (FELIX) at
the FOM-Institute for Plasma Physics in Nieuwegein, The
Netherlands.\cite{oepts95} The molecular-beam machine has been presented
previously \cite{ratsch05,fielicke05a,fielicke07} and only a brief
description is given here. Metal clusters are produced by pulsed laser
ablation of a cobalt rod using the 2$^{\rm nd}$ harmonic output of a Nd:YAG
laser and by subsequent condensation of the plasma in a mixture of Ar in He.
Neutral, anionic, and cationic clusters are produced in this process and pass
through a temperature controllable copper channel. Only cationic
clusters are investigated in this study. Two different experimental
conditions have been used. With a mixture of 0.1\,\% Ar in He at
liquid-nitrogen-temperature, clusters containing six or more cobalt atoms
bind one and, to a much lesser degree, two argon atoms. Under the same conditions
small clusters up to Co$_5^+$ form complexes with up to five argon ligands.
At a higher temperature of 340\,K and using a mixture of 0.3\,\% Ar in He no
attachment to bigger cobalt clusters is observed, whereas Co$_4^+$ and
Co$_5^+$ add one and two rare-gas atoms which is in line with
observations by Minemoto {\em et al.} \cite{minemoto96}.

The molecular beam expands into vacuum and passes through a skimmer and an
aperture before entering the extraction region of a reflectron time-of-flight
mass spectrometer. A counter-propagating pulsed far-IR laser beam delivered
by FELIX is overlapped with the aperture to ensure that all species in the
beam that are detected in the mass spectrometer have been exposed to the IR
radiation. The laser pulse consists of a 1 GHz train of ps-duration
micropulses. The duration of such a macropulse is several $\mu$s while its
energy ranges typically between 15 and 30 mJ, depending on the wavelength.
When the IR radiation is resonant with an IR-allowed transition of the
cluster complex, sequential absorption of single photons can take
place.\cite{oomens03} The resulting heating of the cluster may induce
evaporation of the argon atoms, which leads to depletion of the complexes in
the beam. IR depletion spectra are constructed by recording the ion
intensities of the metal argon complexes as a function of the FELIX
frequency; from these, the far-infrared absorption spectra are obtained as
described before.\cite{fielicke05a} As the detection is mass selective, the
simultaneous measurement of far-IR spectra for different cluster sizes is
possible.

\section{Theory}

The DFT calculations are carried out using the all-electron full-potential code FHI-aims \cite{aims} and the generalized gradient approximation (GGA-PBE) \cite{perdew96} to describe electronic exchange and correlation (xc). For comparison, some calculations are also performed using the local density approximation (PW-LDA) \cite{perdew92}. FHI-aims employs basis sets consisting of atom-centered numerical orbitals. All calculations reported here are done with the ``tier2'' basis set which contains 67 and 45 basis functions for Co and Ar, respectively. As integration grids, we used 158 and 140 radial shells for Co and Ar, respectively, in which the number of integration points is successively decreased from 1202 for the outermost shell to 110 for the innermost one. Convergence tests focused on the D$_{\rm 2d}$ ground-state and the higher-lying D$_{\rm 4h}$ isomer of Co$_4^+$-Ar ({\em vide infra}). Using the hierarchically constructed larger basis sets provided in FHI-aims, as well as denser integration grids, we recomputed the relaxed geometric structure and vibrational spectra of the two isomers, as well as their energetic difference, and found these quantities to be converged within 0.01\,{\AA}, 2 cm$^{-1}$ and 2 meV, respectively, which is fully sufficient for the arguments and conclusions put forward below.

Local structural optimization is done using the Broyden-Fletcher-Goldfarb-Shanno method \cite{recipes}, relaxing all force components to smaller than $10^{-2}$\,eV/{\AA}. After relaxation, the Hessian matrix is determined numerically by finite displacements of all atomic positions by $10^{-3}$\,{\AA}, and then diagonalized to get the vibrational modes. The corresponding infrared-intensities are obtained by taking the derivative of the dipole moments along these modes. In order to facilitate the visual comparison to the experiment, the resulting IR-spectra are finally folded with a gaussian line shape function of half-width 1 $\rm cm^{-1}$.

In order to identify energetically low-lying isomers we tested all structures for the bare clusters that had been discussed previously in the literature. For increasing cluster sizes this approach is likely to miss relevant geometries. This holds for Co in particular, where one has to expect several, differently Jahn-Teller distorted versions of the same structural motif. In addition, we therefore ran extended basin-hopping \cite{wales00} runs to achieve a better sampling of configurational space.
BH runs proceed through a sequence of jumps from one potential-energy surface (PES) minimum to another. For this, an initially random cluster structure is subject to so-called trial moves, which correspond to a random structural modification, followed by a local geometry optimization. The desired importance sampling of energetically lowest-lying isomers and the possibility to surmount barriers on multiple-funnel type PESs is ensured through a Metropolis-type acceptance rule with which a created trial structure replaces the current cluster structure as starting point for the following trial move.

As specific implementation we chose so-called single-particle trial moves, in which a randomly picked atom is displaced in a random direction by a distance of 3.00\,{\AA} (corresponding to 1.5 times the GGA-PBE computed Co dimer bond length). In order to prevent an entropy-driven dissociation of the cluster during the BH run, we disregarded trial moves as well as local relaxations that generate loosely connected or partly dissociated structures characterized by an atom having a nearest-neighbor distance larger than twice the dimer bond length. Similarly discarded were moves that place atoms at distances of less than 0.5\,{\AA} from each other. In view of the specific electronic configuration of Co, we furthermore explicitly included the spin degree of freedom into the sampling procedure, i.e. after the trial atomic displacement the electronic self-consistency cycle was initiated with a random magnetic moment. For specific isomers we further supplemented this by fixed-spin moment calculations enforcing spin-states that had not been found in the BH runs. However, in almost all cases these states turned out to not be local minima of the spin hypersurface, suggesting that the chosen spin sampling in the BH runs yields the relevant spin minima.

Equivalent BH runs were performed for the Ar$_m$-Co$_n^+$ complexes, randomly displacing both Ar and Co atoms. The most favorable structures determined by this completely unbiased treatment of the Ar probe atoms were always Co clusters with Ar adsorbed at a top site, with Ar-Co bond lengths that range from 2.45\,{\AA} for the Co$_4^+$ ground-state to 2.63\,{\AA} for the most stable isomer of Co$_8^+$.

\section{Results and Discussion}

\subsection{Setting the stage}

\begin{figure}[t]
\centering
\includegraphics[width=8.5cm]{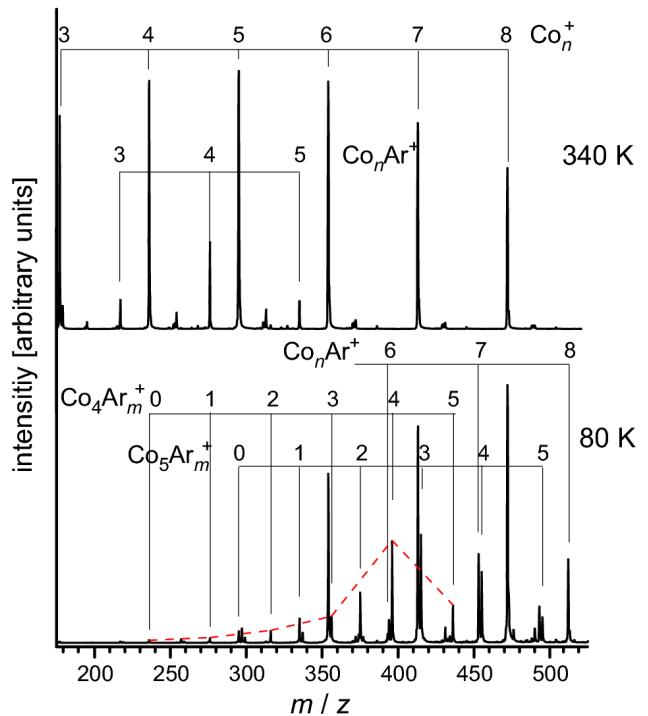}
\caption{Distribution of cationic cobalt clusters and their Ar complexes
obtained at source temperatures of 340 K (top) and 80 K (below). At the
higher temperature rare gas complexes are only observed up to Co$_5^+$,
whereas at low temperature also larger clusters attach Ar and the small
clusters bind multiple Ar atoms. The dashed red line in the lower plot 
connects the maxima of the Co$_4^+$-Ar$_n$ peaks.}
\label{fig1}
\end{figure}

The recorded mass spectra of the cationic cobalt-argon complexes reveal 
profound differences in the binding of the rare-gas ligands to clusters 
containing more or less than six atoms, for both employed complex formation
conditions (77\,K/0.1\%\,Ar vs. 340\,K/0.3\%\,Ar). As shown in Fig.~\ref{fig1}, 
at the low-temperature conditions, in which the bigger clusters
bind only one or at maximum two Ar atoms, Co$_4^+$ and Co$_5^+$ readily
attach up to five ligands. The affinity of these small clusters to the
rare-gas atoms is so pronounced that they still form complexes with one or
two Ar atoms even at a temperature as high as 340\,K. This change of
the interaction with the probe atoms with cluster size is also reflected in
the obtained FIR-MPD spectra, which are measured in the range 75-350
cm$^{-1}$ with no IR absorption bands detected at higher wavenumbers. The
spectra show a clear dependence on the number of adsorbed Ar atoms,
which is strongest for the smallest clusters and is essentially lost for
clusters containing more than seven Co atoms.

Compared to linear absorption spectra, FIR-MPD spectra are known to be 
subject to distortions, resulting from an incomplete fragmentation of complexes 
containing more rare-gas atoms and (cross)anharmonicities in the multiple 
photon excitation process\cite{fielicke05a}. For the first of these effect one 
has to keep in mind that there is no mass selection prior to the fragmentation of a
certain complex. The spectrum of a given complex thus contains both a
contribution from the depletion due to evaporation of the argon ligand as
well as an increase in intensity due to incomplete fragmentation of complexes
containing more argon ligands. This can lead to missing bands, especially at
the low-frequency end of spectra, and is avoided by considering only the
spectra of clusters with the maximum number of argon ligands.
The effect of (cross)harmonicities is to change the width and
positions of existing lines. Both effects have been observed in
preceding work on cationic V$_n^+$ ($n=3-23$) \cite{fielicke04,ratsch05},
Nb$_n^+$ ($n=5-9$) \cite{fielicke05,fielicke07} and Ta$_n^+$ ($n=6-20$)
\cite{gruene07} complexes. However, the influence of the rare-gas atoms was
typically small, with only minor shifts of the frequencies of the absorption
bands on the order of a few cm$^{-1}$ and with the relative intensities of
the observed bands barely affected. This situation is markedly different in
the case of the small cationic Co clusters, where the spectra from different
Ar containing complexes exhibit much more pronounced frequency shifts and
rather strong changes in the relative intensities, cf. Figs. \ref{fig3} and
\ref{fig4} below. This finding together with the indicated unusually strong
interaction with the Ar probe atoms prompts us to use DFT to specifically
investigate the Ar-Co$_n^+$ interaction and its role for the vibrational
spectra, focusing in view of the experimental data on the size range $n = 4
\ldots 8$.

\begin{figure}[t]
\centering
\includegraphics[width=8.5cm]{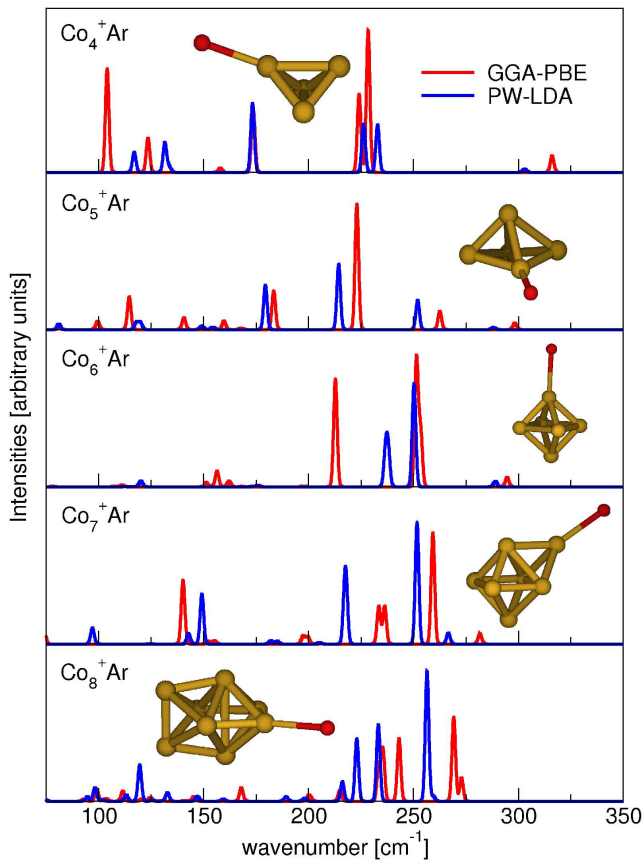}
\caption{Comparison of computed vibrational spectra of the identified most stable isomers of Co$_4^+$-Ar to Co$_8^+$-Ar and using the PW-LDA (blue line) or GGA-PBE (red line) functional. A common scaling factor is applied, matching the main peak of the experimental Co$_6^+$-Ar spectrum. See text and Figs. \ref{fig3}-\ref{fig7} below for a description of the different isomers.}
\label{fig2}
\end{figure}

In general, there is little reason to expect that DFT with present-day local
and semi-local xc functionals yields an accurate description of the intricate
electronic structure of such small Co clusters. The difficulty of electronic
structure calculations to properly account for the delicate balance between
exchange and correlation, between magnetism and chemical bonding, and between
populating $d$ and $s$ orbitals in these systems is already well exemplified
by the still controversial predictions of the true ground-state of the
neutral Co dimer (see e.g. ref. \onlinecite{wang05} and refs. therein). With
respect to vibrational frequencies a common procedure to correct for some
(systematic) errors due to an approximate xc treatment is to apply a general
scaling factor to the calculated frequencies to bring the computed spectra in
better agreement with the measured ones. The limitations of this approach for
the cationic Co clusters is, however, well apparent from Fig. \ref{fig2},
which compiles the computed and scaled vibrational spectra of the identified
most stable isomers of Co$_n^+$-Ar complexes in the targeted size range ({\em
vide infra}), and compares them to the corresponding spectra obtained using
the PW-LDA functional. In light of the particularly controversial situation
for the dimer, we did not follow the usual practice to determine the scaling
factor from the comparison of the computed and experimental dimer vibrational
frequency, but used the main peak of the simple fingerprint spectrum of
Co$_6^+$-Ar, cf. Fig. \ref{fig5} below, instead. However, regardless of which
reference is actually used, the validity of such a common scaling approach
would have shown up in basically identical spectra for all cluster sizes
after scaling the PW-LDA and GGA-PBE data, which is obviously not the case.

In this situation, it makes little sense to strive for a quantitative agreement of the absolute frequencies when comparing the experimental to the theoretical spectra. Additionally considering the mentioned differences between linear IR absorption and FIR-MPD spectra, the best one can possibly hope for is a match between (relative) spacings between different peaks in the spectrum, or, more modestly, to identify some characteristic fingerprint pattern that is unique to a specific isomer and which would then allow for a structural assignment. It is within this perspective that we augment the detailed presentation of the low-lying isomers of Co$_4^+$ to Co$_8^+$ identified in the BH runs with a comparison of the computed and measured vibrational spectra of the corresponding Ar complexes. Despite the discussed limitations, we apply thereby a global scaling to the theoretical spectra simply to enable a better visual comparison. Using the main peak of the experimental Co$_6
 ^+$-Ar spectrum as (to some degree arbitrary) reference, this yields a global scaling factor of 0.881 for GGA-PBE and 0.749 for PW-LDA, respectively.

\subsection{Low-lying isomers and vibrational spectra of their Ar complexes}

\subsubsection{Co$_4^+$}

Our calculations for Co$_4^+$ reveal  as most stable isomer structure a
distorted tetrahedron with D$_{\rm 2d}$ symmetry and a magnetic moment of
7\,$\mu_{\rm B}$. This still quite symmetric geometry has two (opposite)
bonds of 2.50\,{\AA} length, while the remaining four bonds are
2.17\,{\AA}. The next lowest structure is already 0.32\,eV higher
in energy, has a magnetic moment of 9\,$\mu_{\rm B}$ and corresponds to an
even further distorted tetrahedron, in which one bond is elongated to
2.90\,{\AA} to yield a kind of butterfly geometry. Even further up in energy
(0.53\,eV) is the lowest-energy planar structure, with 9\,$\mu_{\rm B}$, a
D$_{\rm 2h}$ symmetry and bond lengths of 2.22\,{\AA} and 2.38\,{\AA}. The
thus obtained energetic order is consistent with the interpretation of
photodissociation spectroscopy data favoring a three-dimensional structure
for the cationic Co tetramer \cite{minemoto96}. A corresponding preference
for a pyramidal ground-state was also obtained in earlier DFT calculations
for neutral Co$_4$ clusters, albeit with a much smaller energetic
difference to the lowest lying planar isomer.\cite{jamorski97,datta07} Theory
hence predicts similar geometries for the neutral and cationic Co$_4$
ground-state, which is in line with a corresponding interpretation derived
from measured bond energy data \cite{hales94}.

\begin{figure}[t]
\centering
\includegraphics[width=8.5cm]{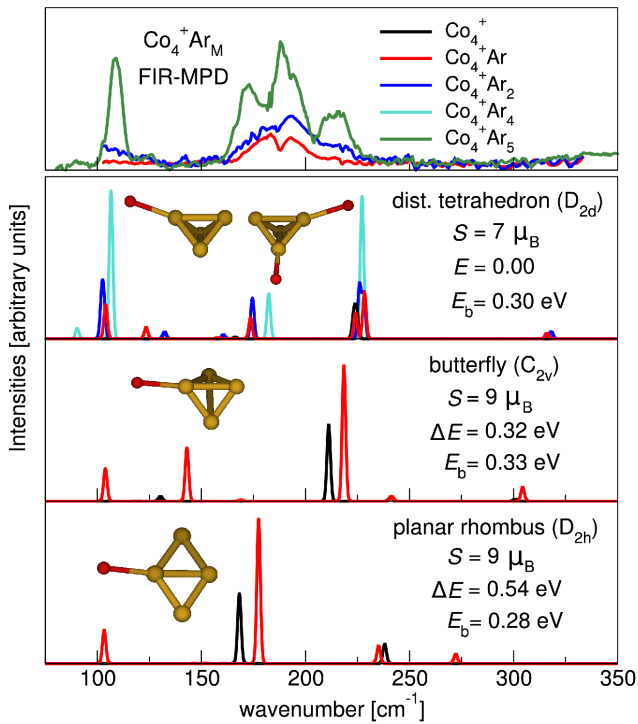}
\caption{Comparison of experimental FIR-MPD spectra (upper panel) with the computed IR-absorption spectra of Ar complexes of the identified three energetically lowest isomers of Co$_4^+$: The most stable distorted tetrahedron (second panel from top), a butterfly-type tetrahedron (third panel from top), and a planar rhombus (bottom panel), see text. Each of the theoretical panels additionally displays a schematic representation of the cluster geometry also indicating the position of the adsorbed Ar atom in the Co$_4^+$-Ar complex, the point group symmetry, the total energy difference $\Delta E$ of the bare cluster with respect to the most stable isomer, and the Ar binding energy $E_{\rm b}$ in the Co$_4^+$-Ar complex. For the lowest-energy isomer, we additionally show a representation of the cluster geometry indicating the position of both adsorbed Ar atoms in the Co$_4^+$-Ar$_2$ complex. In the Co$_4^+$-Ar$_4$ complex each of the four Co atoms is directly coordinated to one Ar atom.}
\label{fig3}
\end{figure}

From this quite consistent picture of the structural motif of the Co$_4^+$
ground-state it is somewhat unexpected to achieve only a rather poor
agreement of the computed vibrational spectra of its Ar complexes with the
corresponding FIR-MPD measurements. Turning first to the experimental data
shown in Fig.~\ref{fig3}, pronounced differences are observed for the spectra
derived from Co$_4^+$-Ar, Co$_4^+$-Ar$_2$ and Co$_4^+$-Ar$_5$. This concerns
frequency shifts and changes in the relative intensities of a group of
presumably three modes in the range 160-210 cm$^{-1}$, as well as the
appearance of a strong mode at 110 cm$^{-1}$ and tentatively a small mode at
90 cm$^{-1}$ only in the spectrum from the complex with five Ar atoms. 
Such apparent missing peaks at the low-frequency end of
the spectrum from complexes can be rationalized by the a Ar-cluster binding
energy for the first ligands and the comparably low photon energy in the 
far-IR, leading to an incomplete fragmentation \cite{fielicke05a}. The frequency
shifts particularly of the central peak observed for the group at higher
wavenumbers are, however, much stronger than those in preceding work on
cationic clusters from group 5 of the Periodic Table
\cite{fielicke04,ratsch05,fielicke05,fielicke07,gruene07}. In this situation,
we note that the spectrum of the Co$_4^+$-Ar$_5$ complex can not be affected
by an incomplete fragmentation of complexes containing even more Ar ligands. 
From the relative abundances of Co$_4^+$+Ar$_4$ and Co$_4^+$+Ar$_5$ in the mass 
spectrum shown in Fig.~\ref{fig2} we expect that the latter complex is
is in fact best described as Co$_4^+$-Ar$_4$ with an additional
rather loosely bound Ar atom. Correspondingly, the spectrum derived
from Co$_4^+$-Ar$_5$ should give a most faithful representation of the vibrational
frequencies of a Co$_4^+$-Ar$_4$ complex.

The computed IR absorption spectra of the most stable D$_{\rm 2d}$ isomer complexed with Ar atoms reproduce the strong low-frequency mode at 120 cm$^{-1}$ quite well, cf. Fig.~\ref{fig3}, supporting the interpretation that the 
absence of this mode in the measured Co$_4^+$+Ar and Co$_4^+$+Ar$_2$ spectra results from incomplete fragmentation. Comparing to the also shown computed spectrum of the bare cluster it is furthermore clear that this agreement is only achieved because of the explicit consideration of the rare-gas atom(s) in the calculations. Only the unexpectedly strong binding of the Ar atom of 0.3\,eV breaks the symmetry of the bare cluster and leads to the appearance of IR-active modes in the low-frequency range at all. This unusual strength of the Ar-Co$_4^+$ interaction prevails for up to four ligands, each time filling one of the top sites at the four corners of the pyramid. Presumably due to the lacking dispersive interactions in the semi-local xc functional, we did, however, not achieve to reliably bind a fifth Ar atom to the cluster at the GGA-PBE level. This supports the interpretation of the measured Co$_4^+$-Ar$_5$ spectrum in terms of the vibrational frequencies of a Co$_4^+$-Ar$_4$ complex, and we correspondingly show in Fig.~\ref{fig3} the computed spectrum of the latter for comparison. Unfortunately, the favorable experiment-theory comparison for the lowest frequency modes does not carry over to the higher-frequency group of bands in the experimental data. In contrast to the more or less evenly spaced three main peaks in this group, theory predicts a large spacing between one lower-energy peak and a higher-energy doublet. While the computed frequencies correspond quite well to the lower and higher frequency main peak of the experimental group, either the central and dominant experimental peak is completely missing or the splitting of the doublet is significantly underestimated in the theoretical spectrum.

The calculated spectra of the second and third lowest-energy isomer also shown in Fig.~\ref{fig3} can neither achieve a better agreement with the experimental data, so that a significant population of a higher-energy isomer is not readily invoked as reason for the discrepancy. At present we are thus unable to provide a clear explanation for the discomforting disagreement between measured and computed data. Two possible reasons could be either an insufficient description provided by the employed GGA-PBE functional or strong (cross-)anharmonicities in the experimental data, which could both particularly affect the splitting of the higher-energy doublet in the D$_{\rm 2d}$ spectrum. The latter is very sensitive to the distortions of the pyramidal ground-state structure caused by Jahn-Teller effects \cite{jamorski97} and the Ar bonding, both of which might not be sufficiently treated at the level of a semi-local xc functional. Simultaneously, one has to keep in mind, however, that the unusually strong Ar bonding in the complex that we will further discuss in the next section severely enhances the multiple photon excitation aspect of FIR-MPD, and makes the comparison to computed linear IR absorption spectra at least for the complexes containing only one or two Ar atoms less justified.

\subsubsection{Co$_5^+$}

For Co$_5^+$ we determine as most  stable isomers three slightly differently
distorted versions of a tetragonal pyramid that are all within a 0.06\,eV
energy range and possess all a magnetic moment of $10 \mu_{\rm B}$. The
energetically highest of the three has still C$_{\rm 2v}$ symmetry, while the
other two are further Jahn-Teller distorted to C$_{\rm 1}$ symmetry by tilts
of the apex Co atom away from the symmetry axis of the slightly folded
rhombohedral base. In all three structures there are sets of shorter Co bonds
of around $\sim 2.2 - 2.3$\,{\AA} and more elongated bonds to the apex atom
of the order of 2.5\,{\AA}. A more symmetric C$_{\rm 4v}$ tetragonal pyramid
with $12 \mu_{\rm B}$ is found at 0.15\,eV above the ground-state, followed
by the lowest-lying planar structure already at 0.81\,eV. This strong
preference for a distorted three-dimensional structure in a high-spin state
is again similar to corresponding findings for the neutral pentamer
\cite{jamorski97,datta07}, further confirming that neutral and cationic Co
clusters exhibit similar geometries \cite{hales94}.

\begin{figure}[t]
\centering
\includegraphics[width=8.5cm]{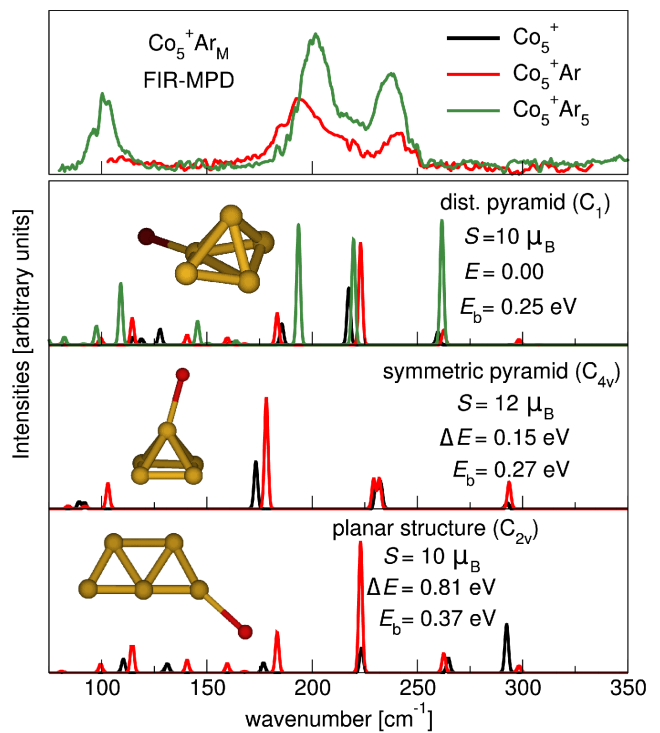}
\caption{Comparison of experimental FIR-MPD spectra (upper panel) with the computed IR-absorption spectra of Ar complexes of identified energetically lowest isomers of Co$_5^+$: One of the three almost degenerate distorted tetragonal pyramids (second panel from top), a more symmetric C$_{\rm 4v}$ pyramid (third panel from top), and a planar structure (bottom panel), see text. Each of the theoretical panels additionally displays a schematic representation of the cluster geometry also indicating the position of the adsorbed Ar atom in the Co$_5^+$-Ar complex, the point group symmetry, the total energy difference $\Delta E$ of the bare cluster with respect to the most stable isomer, and the Ar binding energy $E_{\rm b}$ in the Co$_5^+$-Ar complex. In the Co$_5^+$-Ar$_5$ complex each of the five Co atoms is directly coordinated to one Ar atom.}
\label{fig4}
\end{figure}

The comparison of the computed  IR absorption spectra to the FIR-MPD data is
this time slightly more favorable. Figure \ref{fig4} shows the corresponding
data for one of the degenerate distorted tetragonal pyramids (the spectra of
the other two differ only insignificantly), as well as for the more symmetric
tetragonal pyramid and the planar structure. The measured spectra from
Co$_5^+$-Ar and Co$_5^+$-Ar$_5$ complexes show comparable differences as in
the case of the tetramer complexes: A peak at the low-frequency end of the
spectrum is only observed for the complex containing more Ar atoms and a
group of bands in the range 175-230 cm$^{-1}$ exhibits pronounced frequency
shifts and intensity changes. From the overall shape one would expect at
least three components behind this group and this indeed is what is
found in this frequency range in the computed spectrum for the lowest-energy
distorted pyramids. However, even though only containing two main peaks in
this frequency range, the overall spectrum of the symmetric tetragonal
pyramid isomer does also fit the experimental data rather well, so that we
cannot safely discriminate between these two (in any case rather similar)
structures. What is instead quite clear from Fig.~\ref{fig4} is that the
planar structure can not be reconciled with the measurements.

Due to the already severely distorted geometry of the bare ground-state the influence of the rare-gas atom on the computed spectrum is less dramatic than in the case of the Co tetramer. Rather than leading to the appearance of a series of additional IR-active bands, its effect is more to shift existing peaks and change their intensities. With an again rather strong computed binding energy of around 0.3\,eV this concerns primarily those modes that involve displacements in the direction of the Ar adsorbate. A particularly pronounced example of such a change in intensity is the computed mode at 260 cm$^{-1}$ for the spectra from the Co$_5^+$-Ar and Co$_5^+$-Ar$_5$ complexes, in nice analogy to the variations observed in the corresponding experimental spectra.

\subsubsection{Co$_6^+$}

The two most stable determined isomers of Co$_6^+$ correspond  to slightly
distorted tetragonal bipyramids, with the lowest-energy structure exhibiting
a magnetic moment of $15 \mu_{\rm B}$ and another one 0.10\,eV higher with
$13 \mu_{\rm B}$. Both structures possess D$_{\rm 3d}$ symmetry and bond
lengths around 2.3\,{\AA}. The third lowest-energy geometry is a capped
trigonal bipyramid also with $13 \mu_{\rm B}$, but with 0.56\,eV above the
ground-state this is already significantly higher in energy. With the DFT
calculations for Co$_6$ by Datta {\em et al.} \cite{datta07} also arriving at
a tetragonal bipyramid ground-state, the similarity between the structural
motif of cationic and neutral clusters thus continues also for the hexamer.

\begin{figure}[t]
\centering
\includegraphics[width=8.5cm]{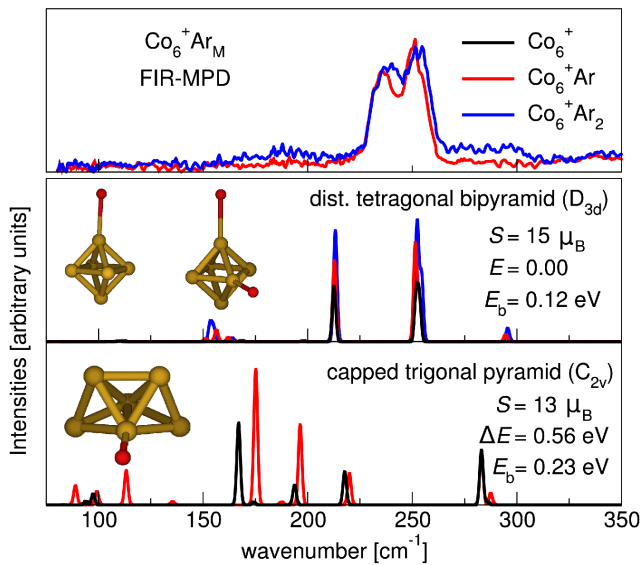}
\caption{Comparison of experimental FIR-MPD spectra (upper panel) with the computed IR-absorption spectra of Ar complexes of identified energetically lowest isomers of Co$_6^+$: The tetragonal bipyramid with $15 \mu_{\rm B}$ (second panel from top) and the capped trigonal bipyramid (third panel from top). The spectrum for a second tetragonal bipyramid with $13 \mu_{\rm B}$ is very similar to the one of the $15 \mu_{\rm B}$ pyramid and omitted for brevity, see text. Each of the theoretical panels additionally displays a schematic representation of the cluster geometry also indicating the position of the adsorbed Ar atom in the Co$_6^+$-Ar complex, the point group symmetry, the total energy difference $\Delta E$ of the bare cluster with respect to the most stable isomer, and the Ar binding energy $E_{\rm b}$ in the Co$_6^+$-Ar complex. For the lowest-energy isomer, we additionally show a representation of the cluster geometry indicating the position of both adsorbed Ar atoms in the Co$_6^+$-Ar$_2$ complex.}
\label{fig5}
\end{figure}

The clear energetic preference for the tetragonal bipyramid motif is
supported by the comparison of computed and experimental vibrational spectra.
The FIR-MPD data is shown in Fig.~\ref{fig5} and exhibits this time
only little differences between the measurements of Co$_6^+$-Ar and
Co$_6^+$-Ar$_2$ complexes. In both cases the spectrum is simple and consists
only of two peaks around 250 cm$^{-1}$. This pattern is fully reproduced in
the calculated spectra for the two tetragonal bipyramids and certainly
inconsistent with the computed spectrum for the capped trigonal bipyramid. In
the case of the cationic hexamer we therefore achieve an unambiguous
assignment of the structural motif. Furthermore, the theoretical spectra for
the identified ground-state structure either bare or with one or two Ar atoms
shown in Fig.~\ref{fig5} are virtually identical. This is distinctly
different to the findings for the smaller Co$_n^+$ clusters, and goes hand in
hand with a significant reduction of the Ar bond strength, computed as
0.12\,eV for the $15 \mu_{\rm B}$ ground-state tetragonal bipyramid.

\subsubsection{Co$_7^+$}

In the case of Co$_7^+$ we identify a capped tetragonal bipyramid with C$_{\rm 3v}$ symmetry and magnetic moment of $16 \mu_{\rm B}$ as most stable structure. Again, we arrive therewith at the same structural motif for the cationic ground-state as the preceding work for the neutral clusters \cite{datta07}. The bond length in the triangle far side from the apex atom is 2.31\,{\AA} and is thus close to the computed bond length of 2.36\,{\AA} of the uncapped Co$_6^+$ ground-state structure, whereas the triangle near side of the apex atom has elongated bonds of 2.43\,{\AA}. The resulting average bond length of 2.34\,{\AA} for the entire cluster is thus slightly extended compared to the corresponding neutral isomer reported by Datta {\em et al.} \cite{datta07} with an average bond length of 2.29\,{\AA}. Energetically only insignificantly higher than the capped tetragonal bipyramid are two spin-variants of a distorted pentagonal bipyramid with C$_{\rm 2v}$ symmetry, namely one with
  $14 \mu_{\rm B}$ only 0.06\,eV higher and one with $16 \mu_{\rm B}$ only 0.08\,eV higher. The structural variety is further complemented by the energetically next lowest isomer, which corresponds to a capped trigonal prism also in a $16 \mu_{\rm B}$ high-spin state and 0.15\,eV higher than the ground-state. Still energetically quite close this is then followed by another, slightly differently distorted $16 \mu_{\rm B}$ C$_{\rm 3v}$ tetragonal bipyramid (0.18\,eV) and a low-symmetry $16 \mu_{\rm B}$ structure at 0.29\,eV.

\begin{figure}[t]
\centering
\includegraphics[width=8.5cm]{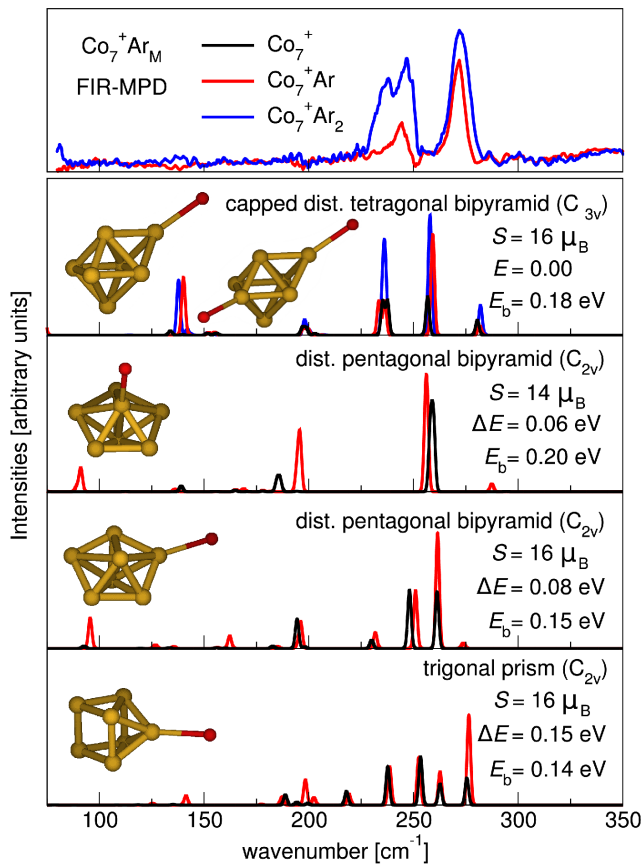}
\caption{Comparison of experimental FIR-MPD spectra (upper panel) with the computed IR-absorption spectra of Ar complexes of identified energetically lowest isomers of Co$_7^+$: The capped tetragonal bipyramid with $16 \mu_{\rm B}$ (second panel from top), two spin-variants of a pentagonal bipyramid (third and fourth panel from top), as well as a capped trigonal prism (bottom panel), see text. Each of the theoretical panels additionally displays a schematic representation of the cluster geometry also indicating the position of the adsorbed Ar atom in the Co$_7^+$-Ar complex, the point group symmetry, the total energy difference $\Delta E$ of the bare cluster with respect to the most stable isomer, and the Ar binding energy $E_{\rm b}$ in the Co$_7^+$-Ar complex. For the lowest-energy isomer, we additionally show a representation of the cluster geometry indicating the position of both adsorbed Ar atoms in the Co$_7^+$-Ar$_2$ complex.}
\label{fig6}
\end{figure}

With this close energetic spacing it is particularly  interesting to compare
to the vibrational spectra from the FIR-MPD experiments and see if the there
determined fingerprint pattern does maybe conform with one of the slightly
higher-energy isomers. The corresponding measurements are shown in Fig.
\ref{fig6}, exhibiting slightly larger differences between the data from the
complexes with different number of Ar atoms than in the case of Co$_6^+$, yet
still significantly less than in the case of the smaller clusters Co$_4^+$
and Co$_5^+$. The computed IR absorption spectrum for the ground-state
tetragonal bipyramid agrees overall rather well with the experiments,
reproducing all three higher-frequency bands. The observed influence of the
Ar atom in the comparison to the bare cluster spectrum is also in line with
the experimental trend, i.e. it is somewhat larger than in the case of
Co$_6^+$, but less than in the case of Co$_4^+$ and Co$_5^+$. Defering the
detailed discussion to the next section we note that this correlates well
with the now again slightly larger computed Ar binding energy of 0.18\,eV.

A major concern in the comparison to the experimental data is the computed strong Ar-induced mode at the lower frequency end of the spectrum. For this, one cannot resort to the argument of apparent missing peaks in the spectra of complexes with one or two Ar atoms, as there are no complexes with more Ar atoms stabilized under the employed beam conditions, cf. Fig.~\ref{fig1}. Comparing the theoretical spectra of the bare and Ar-complexed C$_{\rm 3v}$ ground-state isomer it is, however, clear that the intensity of this particular mode depends sensitively on the interaction with the Ar ligand. The second most stable adsorption site at this isomer is atop one of the basal Co atoms that are on the opposite side from the capping atom, and its binding energy is only 60\,meV lower. In the corresponding spectrum (not shown) the mode does hardly show up. Most likely, we would therefore attribute the discrepancy between experimental and theoretical data with respect to this mode to subtle shortcomings of the employed xc functional in describing the cluster-ligand interaction. Returning to the question of a unique structural assignment, one would probably still conclude that the spectrum of the ground-state capped tetragonal bipyramid agrees overall best with the experimental data. Nevertheless, one also has to recognize that the fingerprint pattern of the different low-energy isomers is not sufficiently distinct to rule out small percentages of these isomers in the experiment as well.

\subsubsection{Co$_8^+$}

The most stable isomer  found for Co$_8^+$ is a double-capped distorted
tetragonal bipyramid with $17 \mu_{\rm B}$, which thus -- like already the
Co$_7^+$ ground-state -- also contains the very stable tetragonal bipyramid
as a structural sub-unit and again coincides with the structural motif
determined as most stable for the corresponding neutral cluster
\cite{datta07}. The geometry has nearly D$_{\rm 2d}$ symmetry, with only bond
distance differences of the order of $\sim 0.01$\,{\AA} reducing it to the
C$_{\rm 2v}$ point group. The next lowest isomer found is only marginally
higher in energy (0.09\,eV) and corresponds to a double-capped trigonal prism
with a magnetic moment of $17 \mu_{\rm B}$. This is followed by another
double-capped tetragonal bipyramid of the same magnetic moment (0.21\,eV),
and then several differently distorted and energetically virtually degenerate
capped pentagonal bipyramid structures at around 0.47\,eV.

\begin{figure}[t]
\centering
\includegraphics[width=8.5cm]{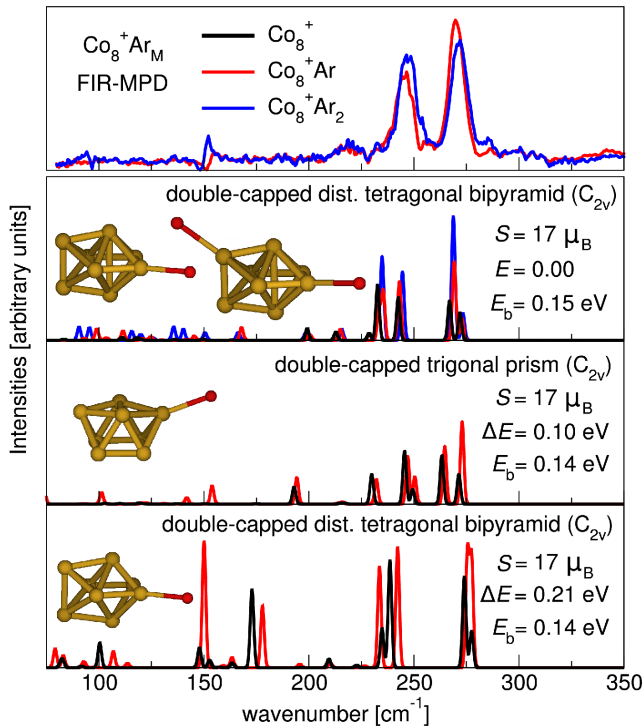}
\caption{Comparison of experimental FIR-MPD spectra (upper panel) with the computed IR-absorption spectra of Ar complexes of identified energetically lowest isomers of Co$_8^+$: The double-capped distorted tetragonal bipyramid with $17 \mu_{\rm B}$ (second panel from top), the double-capped trigonal prism (third panel from top), as well as another double-capped tetragonal bipyramid (bottom panel), see text. Each of the theoretical panels additionally displays a schematic representation of the cluster geometry also indicating the position of the adsorbed Ar atom in the Co$_8^+$-Ar complex, the point group symmetry, the total energy difference $\Delta E$ of the bare cluster with respect to the most stable isomer, and the Ar binding energy $E_{\rm b}$ in the Co$_8^+$-Ar complex. For the lowest-energy isomer, we additionally show a representation of the cluster geometry indicating the position of both adsorbed Ar atoms in the Co$_8^+$-Ar$_2$ complex.}
\label{fig7}
\end{figure}

For this largest studied cluster  the comparison of the computed and measured
vibrational data is unfortunately again rather inconclusive with respect to a
potential structural assignment. The experimental spectra are shown in Fig.
\ref{fig7} and exhibit primarily two main peaks with some suggested shoulders
at the high-frequency end. There are only marginal differences between the
measurements from the Co$_8^+$-Ar and Co$_8^+$-Ar$_2$ complexes, which is
characteristic for all further FIR-MPD spectra recorded for larger clusters
up to 18 Co atoms (not shown). This minor influence of the adsorbed rare-gas
atom is similarly found in all computed spectra for the three most stable
Co$_8^+$ isomers, concomitant with a moderate Ar binding energy of around
0.15\,eV. In all cases the data obtained is furthermore more or less
similarly consistent with the measured fingerprint pattern, with maybe a
slightly better agreement for the ground-state isomer. Put more modestly
there is at least no compelling evidence that it is not the latter structure
that is predominantly observed, and thus there is neither a reason to
question the energetic order determined at the GGA-PBE level or to invoke the
kinetic stabilization of a significant fraction of higher-energy isomers in
the experiments. Part of the reason for the disappointing inconclusiveness
are the rather strong Jahn-Teller deformations found for the high-spin
isomers, which make it partly a matter of semantics to distinguish between
e.g. a double-capped distorted trigonal prism or a double-capped distorted
tetragonal bipyramid. Without clearly distinct symmetries the various
geometries exhibit only quantitative differences in the IR-spectra, which in
turn severely limits a technique like FIR-MPD which relies on an indirect
structure determination through characteristic fingerprint patterns.

\subsection{Nature of the Ar-Co$_n^+$ bonding}

While the comparison of computed and measured vibrational spectra in particular of the smallest studied Co clusters leaves a number of open questions and doubts with respect to the detailed interpretation of the FIR-MPD data, the general trend of the influence of the rare-gas atoms with cluster size is reproduced quite well by the calculations. In general, the Ar-induced changes of the IR-spectra become smaller for larger clusters, correlating nicely with an overall reduction of the Ar-Co$_n^+$ bond strength from an ``unusually'' high value of 0.3\,eV for the lowest lying isomers of Co$_4^+$ to the more ``intuitive'' value of around 0.1\,eV for the largest studied clusters. Particularly for the smallest clusters, the complexation with Ar leads to the appearance of new IR-active modes in the spectra, hand in hand with frequency shifts and splittings of existing peaks and often an increase in the absorption intensity. These effects can be primarily traced back to Ar-induced symmetry breakings in the geometric structure of the cluster, together with an enhanced polarization of the cluster electron density that we will further analyze below. Both effects depend obviously on the strength of the Ar-Co$_n^+$ interaction, rationalizing the correlation with the computed Ar binding energy. The explanation for the observation of a much more pronounced dependence of the FIR-MPD spectra on the number of Ar atoms compared to the preceding work on the cationic group 5 clusters \cite{fielicke04,ratsch05,fielicke05,fielicke07,gruene07} is therefore coupled to the understanding of the larger Ar binding energy at the smaller Co clusters compared to the e.g. $\sim 0.1$\,eV computed also with GGA-PBE for the previously studied $V_n^+$ clusters \cite{fielicke04,ratsch05}.

\begin{figure}[t]
\centering
\includegraphics[width=8cm]{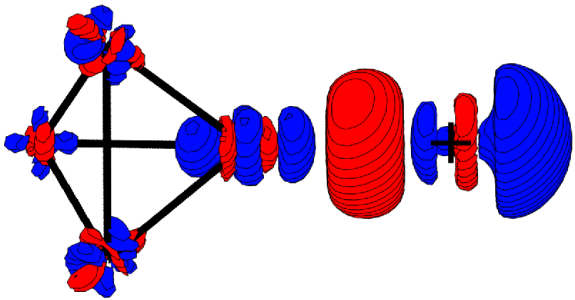}
\caption{Electron density difference induced by the adsorption of  Ar at the
ground-state isomer of Co$_4^+$, see text. For the optimized geometry of the
Co$_4^+$-Ar complex, the plot shows the isosurface corresponding to
0.015\,{\AA}$^{-3}$ electron accumulation (red) and depletion (blue) upon Ar
addition, obtained by subtracting from the full self-consistent electron
density of the Ar-Co$_4^+$ complex the self-consistent electron densities of
the bare Co$_4^+$ cluster and an isolated Ar atom at the corresponding atomic
positions. For better visualization, the drawn skeleton of black ``bonds'' at
the left of the figure represents the cluster geometry, while the black cross
marks the position of the Ar atom.} \label{fig8}
\end{figure}

\begin{figure}[t]
\centering
\includegraphics[width=8.5cm]{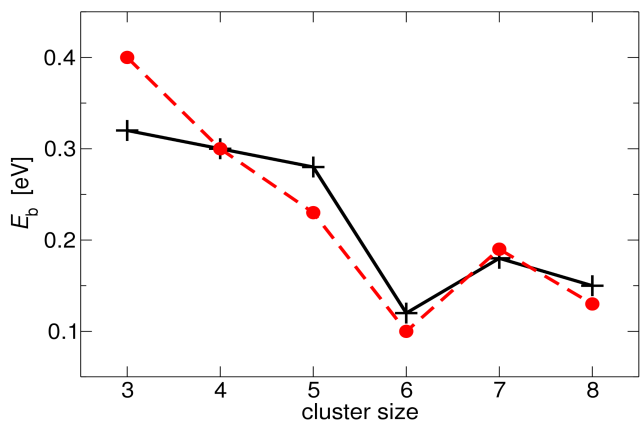}
\caption{Ar binding energy at the most stable adsorption site offered by the
determined ground-state  isomers of Co$_4^+$ to Co$_8^+$. Compared are the
values obtained from the real self-consistent calculation (black crosses) and
the Hartree-type interaction energy obtained when placing Ar into the frozen
electrostatic field formed by the point charges and dipole moments of the
bare cluster (red circles), see text.} \label{fig9}
\end{figure}

As an important key to understand the particularities of the Ar-Co$_n^+$ binding energy, Fig.\ref{fig8} shows a plot of the electron density redistribution when adsorbing an Ar atom at the ground-state isomer of Co$_4^+$. Observable is a strong polarization of the Ar atom, which is equivalently obtained in a corresponding analysis of Ar adsorption at all larger cluster sizes studied. This suggests as the predominant contribution to the Ar-Co$_n^+$ bond a mere polarization of the rare-gas atom in the electrostatic field of the cationic cluster. Further support for this view on the nature of the Ar-Co$_n^+$ interaction comes from a multipole decomposition of the charge density of the determined ground-state isomers of the bare Co$_n^+$ clusters. Via Hirshfeld analysis we compute the point charge and dipole moment on each atom in the bare Co$_n^+$ cluster, yet at the relaxed geometry of the corresponding Co$_n^+$-Ar complex. In a second step we then compute the Hartree-type interaction energy of an Ar atom placed at the position in the Co$_n^+$-Ar complex into the frozen electrostatic field generated by the computed point charges and dipole moments of the bare cluster. The correspondingly obtained binding energies are compared to the fully self-consistently computed Ar binding energies at the different cluster sizes in Fig.~\ref{fig9}. From the rather good agreement we conclude that the polarization of the Ar atom describes indeed the predominant character of the bond, with larger multipole moments and the back-reaction of the Co charge density on the polarized Ar atom contained in the true self-consistent calculation forming only a small correction.

Further disentangling the contributions to the binding energy, we actually compute only a rather small attractive component due to the point charges. With the calculated dipole moments on the Co atoms generally pointing radially outwards of the cluster, a large attraction comes instead from the interaction of the Ar atom with the dipole moment on the Co atom to which it is directly coordinated in the top site position. This is then offset by the mostly repulsive interactions with the less favorably oriented dipole moments on all other Co atoms in the cluster. These findings hold similarly for all of the studied cluster sizes, and even carry over to less stable Ar adsorption sites and other low-energy isomers than the ground-state, for all of which we obtain a similarly good agreement of the Ar binding energy computed in the electrostatic model with the true self-consistent value.

\begin{figure}[t]
\centering
\includegraphics[width=8cm]{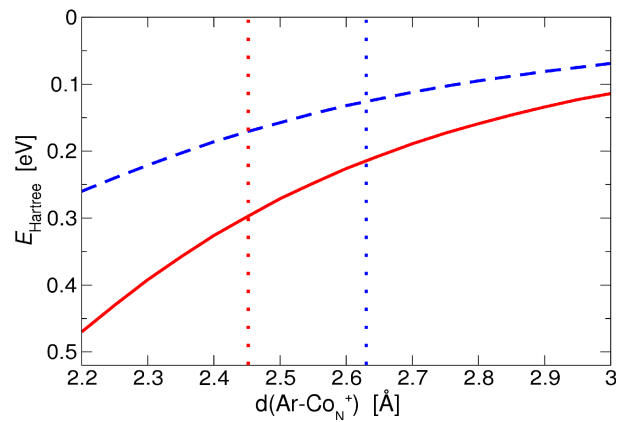}
\caption{Attractive Hartree-type Ar-Co$_n^+$ interaction energy computed in the electrostatic model, see text. Shown is the variation with bond length when moving the Ar atom radially outwards in the Co$_4^+$-Ar (solid line) and in the Co$_8^+$-Ar (dashed line) ground-state complex. The two dotted vertical lines indicate the computed equilibrium bond lengths in the Co$_4^+$-Ar (2.45\,{\AA}) and in the Co$_8^+$-Ar (2.63\,{\AA}) complex.}
\label{fig10}
\end{figure}

On the basis of the thus established electrostatic picture we achieve a rather consistent explanation for the observed trend of the Ar binding energy with Co cluster size, as well as for the stronger interaction compared to the previously studied cationic clusters from group 5 of the Periodic Table. Taking the ground-state isomer of Co$_4^+$ and Co$_8^+$ as examples for a stronger and weaker Ar bonding, respectively, Fig.~\ref{fig10} shows the variation of the interaction energy in the electrostatic model when changing the Ar-Co$_n^+$ bond length by moving the Ar atom radially away from the cluster. Immediately apparent is the smaller attraction in the case of Co$_8^+$, which one can loosely rationalize by the larger number of unfavorably oriented dipole moments at the other Co atoms of the larger cluster to which the Ar atom is not directly coordinated to. Neglecting electronic self-consistency effects, the equilibrium Ar-Co$_n^+$ bond length is determined by the interplay of
  this attractive electrostatic contribution and a strongly repulsive short-range component arising from Pauli repulsion. Since the latter is primarily due to the inner shell electrons of the directly coordinated Co atom, we can assume this component to be rather similar for Co$_4^+$ and Co$_8^+$ to a first approximation. With the stronger electrostatic attraction this then leads to a shorter Ar-Co$_4^+$ bond length than in Ar-Co$_8^+$, and in turn to a stronger Ar binding energy at the smaller cluster.

In the same spirit one can also attribute the comparably stronger interaction of Ar with Co$_n^+$ than with the previously studied V$_n^+$, Nb$_n^+$ and Ta$_n^+$ clusters to this interplay of electrostatic attraction and Pauli repulsion. The distance at which the latter sets in depends of course on the size of the atomic radius, which is smaller for Co than for any of the group 5 elements. With the steep Pauli repulsion setting in at already larger distances, one can in general expect longer equilibrium Ar-metal cluster bond lengths and therewith smaller Ar binding energies for the group 5 elements. As suggested by Fig.~\ref{fig10}, the effect of unfavorably oriented dipole moments is also smaller at such increased interaction distances, so that one would furthermore conclude on a smaller variation of the Ar binding energy with cluster size as in the case of Co. Simplistically equating the Ar bond strength with the influence of the rare-gas atom on the measured FIR-MPD spectra
  as discussed above, the electrostatic picture can thus fully account for the general trends observed in the corresponding experiments on these materials. Tacitly assuming a similar character of the rare-gas$-$metal cluster bond, one would even predict that equally pronounced effects of the probe atoms as in the case of Co will be obtained whenever studying cationic clusters formed of other elements with small atomic radii or for more polarizable heavier rare-gas atoms.

\section{Summary and Conclusions}

Summarizing we have used density-functional theory to study energetically low-lying isomers of cationic Co clusters containing from four to eight atoms. Supplementing our search for structural motifs with first-principles basin hopping sampling the identified ground-state structures agree for each cluster size with the geometry determined in preceding DFT studies for the corresponding neutrals. All structures are in high-spin states and exhibit sizable Jahn-Teller distortions. Our specific motivation has been a detailed comparison of the vibrational properties of the clusters with experimental far-infrared multiple photon dissociation data. While this comparison allows only in few cases for a unique structural assignment and leaves some open questions and doubts in particular for Co$_4^+$, there is at least no compelling evidence that would question the energetic ordering of the isomers obtained at the GGA-PBE level or reciprocally indicate a kinetic stabilization of a significant fraction of higher-energy isomers in the experiments.

Reproduced quite well by the calculations is the general trend of the influence of the rare-gas probe atoms used in the measurements, with Ar-induced changes of the IR-spectra becoming smaller for larger clusters. This correlates nicely with an overall reduction of the calculated Ar-Co$_n^+$ bond strength from 0.3\,eV for the lowest lying isomers of Co$_4^+$ to around 0.1\,eV for the largest studied clusters. This trend can be well explained by the nature of the Ar-Co$_n^+$ interaction, which is predominantly characterized by a polarization of the rare-gas atom in the electrostatic field of the charged cluster. Within the corresponding picture we can also rationalize the less pronounced influence of the probe atoms in preceding FIR-MPD measurements on small cationic clusters formed of group 5 elements, identifying the smaller atomic radius of Co as an important factor.

For the investigated Co$_n^+$ vibrational spectra, complexation with Ar leads in general to the appearance of new IR-active modes, hand in hand with frequency shifts and splittings of existing peaks and often an increase in the absorption intensity. These effects can be primarily traced back to Ar-induced symmetry breakings in the geometric structure of the cluster. The latter become obviously stronger with increasing Ar binding energy, but can never be excluded. For a cluster geometry on the verge of a Jahn-Teller distortion even the tiniest disturbance by a rare-gas ligand may induce symmetry breakings that can then substantially affect the spectrum. In this respect, the present work casts some doubts on the hitherto employed practice of comparing experimental data to computed IR-spectra of bare clusters. The influence of the probe atom can not be judged from a comparison of experimental spectra from different complexes, as the latter reveal only the differences between adsorption of one or more rare-gas atoms, and not with respect to the bare cluster. A computed negligible effect of Ar on the spectrum of one cluster isomer does neither justify to dismiss the ligand in the calculations, since this tells nothing about the liability towards symmetry breaking of other isomers of the same or other cluster sizes. In this situation, explicit consideration of the rare-gas atoms in the modelling is always advisable at little additional computational cost.

\section{Acknowledgements}

This work is supported by the Stichting voor Fundamenteel Onderzoek der Materie (FOM) in providing beam time for FELIX. We thank the FELIX staff for their assistance; in particular Dr. Britta Redlich and Dr. A.F.G. van der Meer. P.G. acknowledges the International Max Planck Research School: \emph{Complex Surfaces in Material Science} for funding. R.G. and K.R. are indebted to Dr. Volker Blum and the FHI-aims development team for useful discussions and technical support.

\end{document}